\documentclass[12pt]{article}
\usepackage{epsf,amsmath}

\begin{document}

\begin{titlepage}

\begin{flushright}
CLNS~02/1791\\
{\tt hep-ph/0207002}\\[0.2cm]
June 28, 2002
\end{flushright}

\vspace{1.0cm}
\begin{center}
\Large\bf 
Subleading Shape Functions and the\\
Determination of \boldmath$|V_{ub}|$\unboldmath
\end{center}

\vspace{1.0cm}
\begin{center}
Matthias Neubert\\[0.1cm]
{\sl F.R. Newman Laboratory for Elementary-Particle Physics\\
Cornell University, Ithaca, NY 14853, USA}
\end{center}

\vspace{1.0cm}
\begin{abstract}
\vspace{0.2cm}\noindent 
It is argued that the dominant subleading shape-function contributions 
to the endpoint region of the charged-lepton energy spectrum in 
$B\to X_u\,l\,\nu$ decays can be related in a model-independent way to
an integral over the $B\to X_s\gamma$ photon spectrum. The square root 
of the fraction of $B\to X_u\,l\,\nu$ events with charged-lepton energy 
above $E_0=2.2$\,GeV can be calculated with a residual theoretical 
uncertainty from subleading shape-function effects that is safely 
below the 10\% level. These effects have therefore a minor impact on 
the determination of $|V_{ub}|$.
\end{abstract}
\vfill

\end{titlepage}

\noindent
{\em 1. Introduction:\/}
One of the most promising strategies for the extraction of the 
Cabibbo--Kobayashi--Maskawa matrix element $|V_{ub}|$ relies on the 
measurement of the inclusive semileptonic $B\to X_u\,l\,\nu$ decay
rate in the endpoint region of the charged-lepton energy spectrum, 
which is inaccessible to decays with a charm hadron in the final
state \cite{Bornheim:2002du}. Non-perturbative effects can be 
controlled systematically by using a twist expansion 
\cite{Neubert:1993ch,Bigi:1993ex} and soft-collinear factorization 
theorems \cite{Korchemsky:1994jb,Bauer:2001yt}. At leading order in 
$1/m_b$, bound-state effects are incorporated by a shape function 
accounting for the ``Fermi motion'' of the $b$ quark inside the $B$ 
meson. This function can be determined experimentally from the photon 
energy spectrum in inclusive radiative $B\to X_s\gamma$ decays
\cite{Neubert:1993ch}.

Recently, there have been first discussions of the structure of
subleading-twist contributions to the $B\to X_s\gamma$ and 
$B\to X_u\,l\,\nu$ spectra, which (at tree level) can be parameterized
in terms of four subleading shape functions \cite{Bauer:2001mh}. The 
phenomenological impact of these functions on the inclusive 
determination of $|V_{ub}|$ has been investigated in
\cite{Bauer:2002yu,Leibovich:2002ys}. These authors point out that
certain $1/m_b$ corrections related to chromo-magnetic interactions 
appear to be enhanced by large numerical coefficients. They conclude
that the ignorance about the functional form of the subleading 
shape functions would lead to a significant theoretical uncertainty in 
the determination of $|V_{ub}|$, which could only be reliably reduced 
if the lower cut on the lepton energy were taken below the region where 
Fermi-motion effects are important (i.e., below 2\,GeV or so). For a 
value $E_0=2.2$\,GeV, as employed in a recent analysis reported by 
the CLEO collaboration \cite{Bornheim:2002du}, the resulting 
uncertainty on $|V_{ub}|$ was estimated to be at the 15\% level 
\cite{Bauer:2002yu}. While using simple models the correction was found 
to be negative, it was argued that the sign of the effect was uncertain 
in general \cite{Leibovich:2002ys}.

In the present note we explore in more detail the origin of the 
``enhanced'' corrections found in these papers. Our main point is that 
the first moments (but not higher moments) of the subleading shape 
functions give a large, non-vanishing contribution to the integral 
over the lepton spectrum even if the lower lepton-energy cut is taken 
out of the endpoint region. This effect corresponds to a calculable 
correction of order $\Lambda_{\rm QCD}^2/(m_b\,\Delta E)$, where
$\Delta E=M_B/2-E_0$. The hadronic uncertainty inherent in the 
modeling of subleading shape functions must therefore be estimated 
with respect to this contribution. When this is done, the remaining 
theoretical uncertainty is found to be much less than what has been 
estimated in \cite{Bauer:2002yu,Leibovich:2002ys}. We show how the 
effect of the first moments of the subleading shape functions can be 
isolated and expressed in a model-independent way in terms of the 
photon energy spectrum measured in $B\to X_s\gamma$ decays. We then 
estimate the numerical effect of the residual higher-twist corrections 
and find their impact on the $|V_{ub}|$ determination to be small, 
safely below the level of 10\%.

\vspace{0.5cm}\noindent
{\em 2. Charged-lepton energy spectrum:}
The quantity of primary interest to the determination of $|V_{ub}|$ is 
the normalized fraction of $B\to X_u\,l\,\nu$ events with 
charged-lepton energy above a threshold $E_0$ chosen so as to 
kinematically suppress the background from $B\to X_c\,l\,\nu$ decays,
\begin{equation}
   F_u(E_0) = \frac{1}{\Gamma(B\to X_u\,l\,\nu)} 
   \int\limits_{E_0}^{M_B/2}\!dE_l\,
   \frac{d\Gamma(B\to X_u\,l\,\nu)}{dE_l} \,.
\end{equation}
When combined with a prediction for the total $B\to X_u\,l\,\nu$ decay
rate, knowledge of the function $F_u(E_0)$ allows one to turn a 
measurement of the branching ratio for $B\to X_u\,l\,\nu$ events with
$E_l>E_0$ into a determination of $|V_{ub}|$. 

In the formal limit where the ``energy window'' $\Delta E=M_B/2-E_0$
is such that $\Lambda_{\rm QCD}\ll\Delta E\ll m_b$, Fermi-motion 
effects can be neglected, and the function $F_u(E_0)$ can be 
calculated using the operator product expansion. At tree level the 
result is
\begin{equation}\label{FuOPE}
   F_u(E_0) = \frac{2(2\Delta E-\bar\Lambda)}{m_b}
   - \frac{\lambda_1+33\lambda_2}{3m_b^2} + O[(\Delta E/m_b)^3] \,,
\end{equation}
where $\bar\Lambda=M_B-m_b$, and $2\Delta E-\bar\Lambda=m_b-2E_0$ is 
twice the width of the energy window in the parton model. The hadronic 
parameters $\lambda_1$ and $\lambda_2$ measure the $b$-quark kinetic 
energy and chromo-magnetic interaction inside the $B$ meson. Note that 
while the leading contribution in (\ref{FuOPE}) is proportional to the 
width of the energy window, the power corrections are independent of 
$\Delta E$. As a result, the relative size of the power corrections 
strongly increases as the energy cut $E_0$ is raised toward the 
kinematic endpoint (corresponding to $\Delta E\to 0$). Although this 
simple analysis breaks down as $\Delta E\sim\bar\Lambda$, it explains 
that the origin of the large power corrections found in 
\cite{Bauer:2002yu,Leibovich:2002ys} is the kinematic suppression of 
the leading-order term. 

For realistic values of the energy threshold the quantity $\Delta E$ 
is of order $\bar\Lambda$, and the operator product expansion must be 
replaced by the twist expansion \cite{Neubert:1993ch,Bigi:1993ex}. At 
subleading order in $1/m_b$ the tree-level expression for $F_u(E_0)$ 
can then be written as 
\begin{equation}\label{Fuint}
   F_u(E_0) = 2 \int\limits_{-\bar\Lambda}^{2\Delta E-\bar\Lambda}\!\!
   d\omega\,\frac{2\Delta E-\bar\Lambda-\omega}{m_b-\omega}\,
   {\cal F}_u(\omega) \,,
\end{equation}
where
\begin{eqnarray}\label{Fudef}
   {\cal F}_u(\omega) 
   &=& f(\omega) + \frac{1}{m_b}\,\left[
    \frac{t(\omega)}{2} - G_2(\omega) - 2\omega\,f(\omega)
    + 3 H_2(\omega) - h_1(\omega) \right] + \dots \nonumber\\
   &\equiv& {\cal F}_s(\omega) + \frac{2}{m_b}\,\Big[
    H_2(\omega) - h_1(\omega) \Big] + \dots
\end{eqnarray}
is a combination of the leading and subleading shape functions 
\cite{Bauer:2002yu}, and the dots denote higher-order terms in the 
expansion. The function ${\cal F}_s(\omega)$ defined by the second 
relation is related to the normalized photon energy spectrum in 
$B\to X_s\gamma$ decays, $S(E_\gamma)$, by (the factor 2 results from 
the Jacobian $d\omega/dE$)
\begin{equation}
   2 {\cal F}_s(m_b-2E_\gamma) = S(E_\gamma)
   \equiv \frac{1}{\Gamma(B\to X_s\gamma)}\, 
   \frac{d\Gamma(B\to X_s\gamma)}{dE_\gamma} \,.
\end{equation}
It is important in this context that the shape of the $B\to X_s\gamma$ 
photon spectrum is largely insensitive to possible effects of New 
Physics \cite{Kagan:1998ym}, so ${\cal F}_s(\omega)$ can be extracted 
from the data in a model-independent way. When we include radiative 
corrections below, $S(E_\gamma)$ will still denote the photon energy 
spectrum, normalized however on an interval 
$E_\gamma^{\rm min}<E_\gamma<M_B/2$, with $E_\gamma^{\rm min}$ 
sufficiently small to be out of the shape-function region.

The combination of subleading shape functions remaining in the last 
line of (\ref{Fudef}) parameterizes chromo-magnetic interactions in
the $B$ meson. The moment expansion of these functions yields 
\cite{Bauer:2001mh}
\begin{equation}
   H_2(\omega) - h_1(\omega)
   = -2\lambda_2\,\delta'(\omega) - \frac{\rho_2}{2}\,\delta''(\omega)
   + \dots \,,
\end{equation}
where $\rho_2$ is a $B$-meson matrix element of a local dimension-6 
operator. In the limit where $\Delta E\gg\bar\Lambda$, only the first 
moment yields a non-zero contribution to the function $F_u(E_0)$ in 
(\ref{Fuint}), because the weight function under the integral is linear 
in $\omega$ (to first order in $1/m_b$). On the other hand, near the 
endpoint of the lepton spectrum all moments of the shape functions become 
equally important \cite{Neubert:1993ch,Bigi:1993ex}. In between these two 
extremes there is a transition region, where only the first few moments 
of the shape functions give significant contributions. Theoretical 
studies of the photon spectrum in $B\to X_s\gamma$ decays have shown that 
this transition region corresponds to values $E_0\sim 2.0$--2.3\,GeV (for 
yet lower values, Fermi-motion effects become unimportant) 
\cite{Kagan:1998ym}. To account for the effect of the first moment we 
define a new subleading shape function
\begin{equation}
   s(\omega) = H_2(\omega) - h_1(\omega) + 2\lambda_2\,f'(\omega) \,,
\end{equation}
whose normalization and first moment vanish, and whose contribution to 
the quantity $F_u(E_0)$ therefore vanishes for $\Delta E\gg\bar\Lambda$. 
Inserting this definition into relation (\ref{Fudef}), and using that 
${\cal F}_s(\omega)=f(\omega)+\dots$ to leading order in $1/m_b$, we 
obtain from (\ref{Fuint})
\begin{equation}
   F_u(E_0) = 2 \int\limits_{-\bar\Lambda}^{2\Delta E-\bar\Lambda}\!\!
   d\omega \left\{ \frac{2\Delta E-\bar\Lambda-\omega}{m_b-\omega}
   \left[ {\cal F}_s(\omega) + \frac{2s(\omega)}{m_b} \right]
   - \frac{4\lambda_2}{m_b^2}\,{\cal F}_s(\omega) \right\} 
   + \dots \,.
\end{equation}
Taking into account the known $O(\alpha_s)$ corrections to the leading 
term in the twist expansion \cite{Leibovich:1999xf,Neubert:2001sk}, 
and rewriting the contribution involving ${\cal F}_s(\omega)$ as a
weighted integral over the normalized photon energy spectrum in 
$B\to X_s\gamma$ decays, we get our final result\footnote{Using an 
integration by parts, this result can be rewritten as a weighted integral 
over the fraction $F_s(E)$ of $B\to X_s\gamma$ events with photon energy 
above $E$, normalized such that $F_s(E_\gamma^{\rm min})=1$.} 
\begin{equation}\label{final}
   F_u(E_0) = \left( 1 + \frac{2\Lambda_{\rm SL}(E_0)}{m_b} \right) 
   \!\int\limits_{E_0}^{M_B/2}\!dE_\gamma\,w(E_\gamma,E_0)\,
   S(E_\gamma) + \dots \,, 
\end{equation}
with the weight function
\begin{equation}\label{weight}
   w(E_\gamma,E_0) = 2 \left( 1 - \frac{E_0}{E_\gamma} \right)
   \left\{ 1 + \frac{\alpha_s(\mu)}{\pi} \left[
   k_{\rm pert}(E_\gamma^{\rm min}) - \frac{10}{9}\,
   \ln\left( 1 - \frac{E_0}{E_\gamma} \right) \right]
   \right\} - \frac{8\lambda_2}{m_b^2} \,,
\end{equation}
and the subleading shape-function contribution
\begin{equation}\label{Lambda}
   \Lambda_{\rm SL}(E_0) = 
   \frac{\int\limits_{-\bar\Lambda}^{2\Delta E-\bar\Lambda}\!\!
         d\omega\,(2\Delta E-\bar\Lambda-\omega)\,s(\omega)}
        {\int\limits_{-\bar\Lambda}^{2\Delta E-\bar\Lambda}\!\!
         d\omega\,(2\Delta E-\bar\Lambda-\omega)\,f(\omega)} \,.
\end{equation}
The factor 2 in front of $\Lambda_{\rm SL}(E_0)$ in (\ref{final}) is 
inserted so that $\Lambda_{\rm SL}(E_0)/m_b$ is the subleading 
shape-function correction to $|V_{ub}|$. We stress that, by definition,
$\Lambda_{\rm SL}(E_0)$ is a parameter of order $\Lambda_{\rm QCD}$ 
that vanishes for $\Delta E\gg\bar\Lambda$. It is thus a true measure 
of shape-function effects. On the contrary, the power corrections 
studied in \cite{Bauer:2002yu,Leibovich:2002ys} arise predominantly 
from the $\lambda_2/m_b^2$ correction to the weight function. 

The expression for the perturbative coefficient $k_{\rm pert}$ in 
(\ref{weight}) can be obtained from the results of 
\cite{Kagan:1998ym,DeFazio:1999sv}. It reads
\begin{equation}
   k_{\rm pert}(E_\gamma^{\rm min})
   = - \frac{35}{9} - \frac23\ln^2\delta - \frac73\ln\delta
   + \sum_{\substack{i,j=2,7,8 \\ i\ge j}}
   \frac{C_i(\mu)\,C_j(\mu)}{\big[C_7(\mu)\big]^2}\,f_{ij}(\delta) \,,
\end{equation}
where $\delta=1-E_\gamma^{\rm min}/\langle E_\gamma\rangle$ depends on 
the lower boundary of the energy interval used to normalize the 
$B\to X_s\gamma$ photon spectrum, $C_i(\mu)$ are leading-order Wilson 
coefficients in the effective weak Hamiltonian for $B\to X_s\gamma$ 
transitions, and the functions $f_{ij}(\delta)$ can be found in 
\cite{Kagan:1998ym}. In the definition of $\delta$ we use the central
value of the CLEO result for the average photon energy above 2\,GeV, 
$\langle E_\gamma\rangle=(2.346\pm 0.034)$\,GeV \cite{Chen:2001fj}, as 
a substitute for $m_b/2$.

\vspace{0.5cm}\noindent
{\em 3. Numerical results:}
The value of the coefficient $k_{\rm pert}$ is sensitive to the choice 
of the renormalization scale $\mu$ and the value of the quark-mass ratio 
$m_c/m_b$ used in the evaluation of charm-quark loops. We take 
$\mu_b=m_b(m_b)=4.2$\,GeV as our central value for the renormalization 
scale and vary $\mu$ between $\mu_b/2$ and $2\mu_b$. We use a running 
charm-quark mass to evaluate the loop functions \cite{Gambino:2001ew}, 
taking $m_c(\mu)/m_b(\mu)=0.23\pm 0.03$. The results for $k_{\rm pert}$ 
corresponding to two different choices of $E_\gamma^{\rm min}$ are
\begin{equation}
   k_{\rm pert} = 
   \begin{cases}
    \phantom{-}0.07\pm 0.10 \,;
     & E_\gamma^{\rm min}=1.5\,\text{GeV} \,, \\
    -0.20\pm 0.08 \,;
     & E_\gamma^{\rm min}=1.75\,\text{GeV} \,.
   \end{cases}
\end{equation}
Since the effect of this correction is very small, one should not 
consider the small variation of $k_{\rm pert}$ as a measure of the
perturbative uncertainty in the weight function (\ref{weight}). 
Typically, we expect $O(\alpha_s^2)$ corrections to contribute at the 
level of 5\% of the tree-level term. A corresponding uncertainty will 
be included in our numerical analysis below.

The power correction to the weight function in (\ref{weight}) may be
rewritten as
\begin{equation}
   \frac{8\lambda_2}{m_b^2} \approx 
   \frac{m_{B^*}^2-m_B^2}{2\langle E_\gamma\rangle^2}
   \approx 0.044 \,.
\end{equation}
Alternatively, using $\lambda_2=(0.12\pm 0.02)$\,GeV$^2$ and
$m_b=(4.72\pm 0.06)$\,GeV we obtain the value $0.043\pm 0.007$, which 
will be used in our numerical analysis. The size of this correction is 
not anomalously large; however, its impact is significant because it 
competes with terms proportional to the small difference 
$(1-E_0/E_\gamma)$. In the endpoint region this difference scales like 
$\Lambda_{\rm QCD}/m_b$, and so the $\lambda_2/m_b^2$ term is of 
relative order $1/m_b$.

Our final focus is on the subleading shape-function contribution 
$\Lambda_{\rm SL}(E_0)$ defined in (\ref{Lambda}). Little is known 
about the function $s(\omega)$ except that its normalization and first 
moment vanish, and that its second moment, $M_2^{(s)}=-\rho_2$, is 
given by a hadronic matrix element expected to be of order 
$(0.5\,\text{GeV})^3$ with undetermined sign. As a result, the 
functional form and sign of $\Lambda_{\rm SL}(E_0)$ cannot be predicted 
at present. However, the fact that $\Lambda_{\rm SL}(E_0)$ must 
approach zero as $E_0$ is lowered to a value of about 2\,GeV (below 
which shape-function effects from higher moments are irrelevant) 
ensures that its impact on the determination of $|V_{ub}|$ is small. 
To substantiate this claim we investigate several models for the 
subleading shape function in more detail. For the leading-order 
function we take the ansatz \cite{Kagan:1998ym}
\begin{equation}
   f(\omega) = \frac{1}{\bar\Lambda}\,g_a(x) \,, \qquad
   \text{with~} x = 1 + \frac{\omega}{\bar\Lambda} \ge 0 \,,
\end{equation}
where $g_a(x)=[a^a/\Gamma(a)]\,x^{a-1}\,e^{-a x}$. The parameter 
$a$ must be larger than 1 and is fixed so that the second moment of 
$f(\omega)$ equals $-\lambda_1/3$ \cite{Neubert:1993ch}, yielding 
$a=-3\bar\Lambda^2/\lambda_1$. We assume that the subleading function
$s(\omega)$ is finite everywhere in the interval 
$-\bar\Lambda\le\omega<\infty$, but we do not require that this 
function vanish at the endpoint.

\begin{figure}[t]
\epsfxsize=14.0cm
\centerline{\epsffile{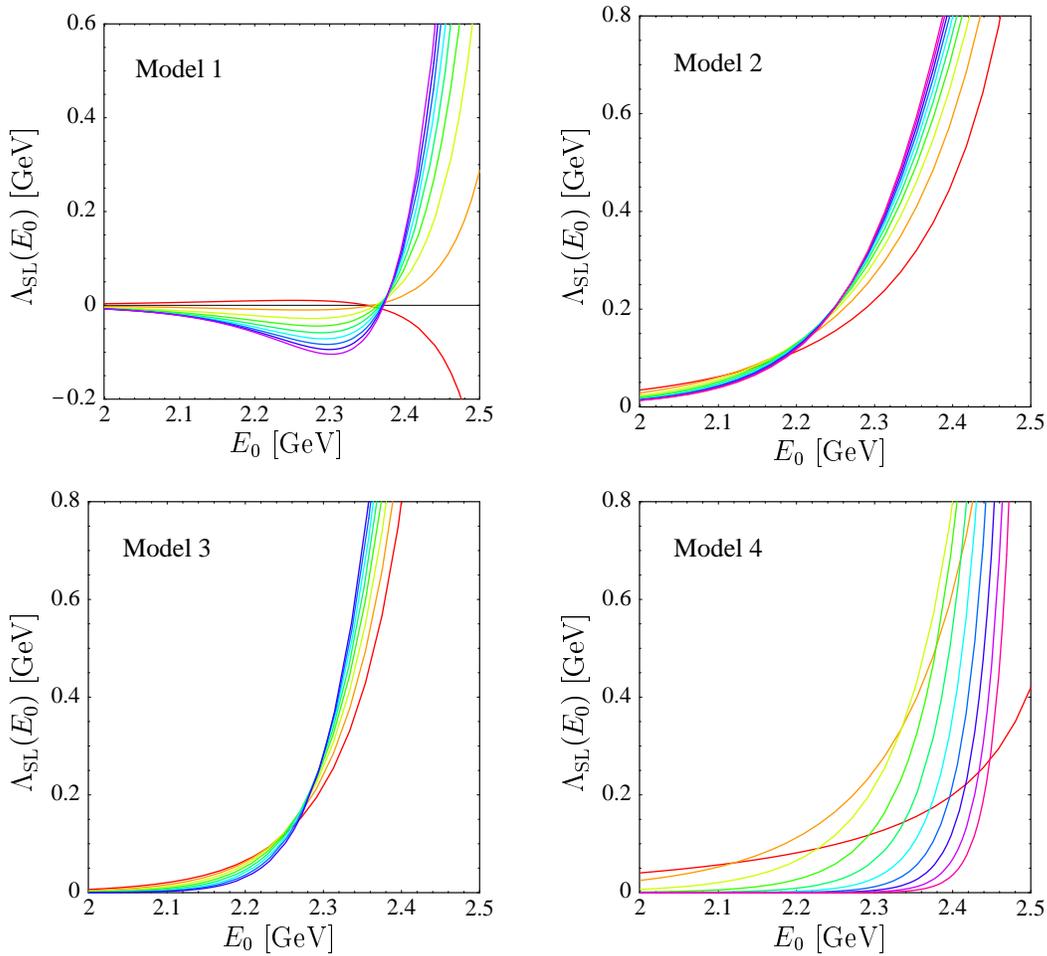}}
\vspace{0.2cm}
\centerline{\parbox{15cm}{\caption{\label{fig:subl}
Model predictions for the subleading shape-function correction
$\Lambda_{\rm SL}(E_0)$ as a function of the cut $E_0$. For each 
model, the parameter $b$ is varied between the minimal allowed value 
(red) and 10 (blue) in steps of 1. The sign of $\Lambda_{\rm SL}(E_0)$ 
is undetermined.}}}
\end{figure}

The model functions adopted in \cite{Bauer:2002yu} are such that 
$s(\omega)$ is set to zero, and so $\Lambda_{\rm SL}(E_0)$ vanishes by 
construction. The model functions used in \cite{Leibovich:2002ys} 
correspond to the ansatz
\begin{equation}
   s(\omega) = \frac{2\lambda_2}{\bar\Lambda^2}\,
    \Big[ g_a'(x) - g_b'(x) \Big] \quad \text{with~}  b\ge 2 \,, 
   \qquad \text{(model 1)}
\end{equation}
where the lower bound on the parameter $b$ is enforced by the 
requirements that $s(\omega)$ be finite at the endpoint 
$\omega=-\bar\Lambda$ and have vanishing normalization and first 
moment. A property of this model is that also the second moment of 
$s(\omega)$ vanishes. Three alternative choices for $s(\omega)$ with 
non-zero second moment $M_2^{(s)}$ are
\begin{equation}
   s(\omega) = \frac{M_2^{(s)}}{2\bar\Lambda^3}\,
   \begin{cases}
    \displaystyle \frac{2ab}{a-b}\,\Big[ g_b(x) - g_a(x) \Big] \quad
    \text{with~} b\ge 1 \,,
     & \text{(model 2)} \\
    \displaystyle \phantom{\bigg|} g_b''(x) \quad \text{with~}
     b\ge 3 \,, & \text{(model 3)} \\
    \displaystyle b^3\,e^{-b x} \left( 1 - 2b x + \frac{b^2 x^2}{2}
     \right) \quad \text{with~} b>0 \,. \quad 
     & \text{(model 4)}
   \end{cases}
\end{equation}
Figure~\ref{fig:subl} shows results for $\Lambda_{\rm SL}(E_0)$ obtained
in the various models, using $\bar\Lambda=0.5$\,GeV, 
$\lambda_1=-0.3$\,GeV$^2$, and $M_2^{(s)}=(0.5\,\text{GeV})^3$ as input 
parameters, and varying the parameter $b$ over a wide range of values. 
Although the details of the subleading shape function $s(\omega)$ are 
rather different in the four cases, all models exhibit the same general 
features. While $\Lambda_{\rm SL}(E_0)$ can be large close to the 
kinematic endpoint, it takes values of order $\bar\Lambda$ for 
$E_0\sim 2.35$\,GeV and quickly decreases as $E_0$ is lowered below
2.3\,GeV. For $E_0=2.2$\,GeV we find values of $\Lambda_{\rm SL}(E_0)$ 
of at most 130\,MeV (model~2), corresponding to a power correction to 
the extraction of $|V_{ub}|$ of less than 3\%. Although our choice of 
model functions is meant as an illustration only, we believe the rapid 
decrease of $\Lambda_{\rm SL}(E_0)$ for $E_0<2.3$\,GeV is a general 
result. It appears to be extremely unlikely that with a reasonable 
shape of $s(\omega)$ and a natural size of the second moment $M_2^{(s)}$ 
the power correction $\Lambda_{\rm SL}(E_0)/m_b$ could be as large as 
10\%.

\vspace{0.5cm}\noindent
{\em 4. Conclusion:}
In summary, we have studied the impact of subleading shape functions
on the determination of $|V_{ub}|$ from the combination of weighted
integrals over energy spectra in inclusive $B\to X_u\,l\,\nu$ and 
$B\to X_s\gamma$ decays. We have argued that for a lower energy cut 
$E_0=2.2$\,GeV as employed in a recent CLEO analysis one is in a 
transition region, where Fermi-motion effects are dominated by the 
first few moments of the leading and subleading shape functions. The 
dominant power correction (the only one that remains when the cut is 
lowered below about 2\,GeV) results from the first moment of the 
subleading shape function, which is known in terms of the hadronic 
parameter $\lambda_2$.

\begin{table}
\caption{\label{tab:Fu}
Illustrative theoretical predictions for the fraction $F_u(E_0)$ of 
$B\to X_u\,l\,\nu$ events with charged-lepton energy $E_l>E_0$, 
assuming a perfect measurement of the $B\to X_s\gamma$ photon 
spectrum (see text for explanation).}
\vspace{0.2cm}
\begin{center}
\begin{tabular}{||c|ccc|c|c||} 
\hline\hline
$E_0$ [GeV] & LO & NLO & $1/m_b$ & total & residual error \\ 
\hline
2.0 & 0.271 & $0.041\pm 0.014$ & $-0.040\pm 0.006$
 & $0.273\pm 0.015$ & $\pm 0.003$ \\ 
2.1 & 0.195 & $0.033\pm 0.010$ & $-0.037\pm 0.006$
 & $0.191\pm 0.011$ & $\pm 0.005$ \\ 
2.2 & 0.126 & $0.024\pm 0.006$ & $-0.033\pm 0.005$
 & $0.117\pm 0.008$ & $\pm 0.006$ \\ 
2.3 & 0.068 & $0.015\pm 0.004$ & $-0.026\pm 0.004$
 & $0.057\pm 0.006$ & $\pm 0.008$ \\ 
\hline\hline
\end{tabular}
\end{center}
\end{table}

Our main result is given in (\ref{final}) and (\ref{weight}). To exhibit 
its features, let us assume that a perfect measurement of the 
$B\to X_s\gamma$ photon spectrum is available in the energy range above 
$E_\gamma^{\rm min}=1.5$\,GeV. (For the purpose of illustration, we 
use a fit to the CLEO data in \cite{Chen:2001fj}.) We then calculate the 
fraction of $B\to X_u\,l\,\nu$ events with charged-lepton energy above 
$E_0$ for different values of the cut. The results are summarized in 
Table~\ref{tab:Fu}. Columns 2, 3, and 4 show the contributions from the 
tree-level term, the $O(\alpha_s)$ corrections, and the power correction 
to the weight function in (\ref{weight}), including theoretical 
uncertainties from input parameter variations as detailed above. The 
next column shows the total result, while the final column gives an 
estimate of the residual uncertainty from subleading shape-function 
effects, as parameterized by the term $2\Lambda_{\rm SL}(E_0)/m_b$ in 
(\ref{final}). We show the largest uncertainty obtained in the four 
classes of models considered earlier. We observe that the power 
correction to the weight function has a significant impact, which as 
anticipated is by far the dominant effect of subleading shape functions. 
For $E_0=2.2$\,GeV, the power correction leads to a reduction of the 
predicted value for $F_u(E_0)$ by $(26\pm 6)\%$, corresponding to a 
13\% enhancement of the extracted value of $|V_{ub}|$. This is in good 
agreement with the estimate given in \cite{Bauer:2002yu}. 

The most important implication of our analysis is that subleading
shape-function effects do not entail a significant limitation on the
extraction of $|V_{ub}|$. This assessment differs from the conclusion
reached in \cite{Bauer:2002yu,Leibovich:2002ys}, where is was argued
that these effects could not be controlled reliably unless the cut 
$E_0$ could be lowered outside the shape-function region. The new 
element of our analysis is that we identify the first moment of the 
subleading shape-function as the dominant source of power corrections 
and show how its contribution can be expressed in terms of an integral 
over the $B\to X_s\gamma$ photon spectrum. We have estimated the 
residual uncertainty on $|V_{ub}|$ from subleading shape-function 
effects by using four different classes of model functions and found 
corrections of at most 3\% (with $E_0=2.2$\,GeV). The smallness of 
this effect can be understood on the basis that it is a power 
correction of the form $\Lambda_{\rm SL}(E_0)/m_b$ with a hadronic 
parameter $\Lambda_{\rm SL}(E_0)=O(\Lambda_{\rm QCD})$ that vanishes 
as $E_0$ is lowered below about 2\,GeV. We thus conclude that, very 
conservatively, the residual uncertainty on $|V_{ub}|$ is less than 
10\%.

The main result of this letter is the new expression for the weight
function in (\ref{weight}), which now includes the leading power 
correction. Perhaps the largest uncertainty in this method for 
determining $|V_{ub}|$ is due to (largely unknown) corrections from 
violations of quark--hadron duality, and 
from spectator-dependent effects such as weak annihilation and Pauli
interference \cite{Bigi:1993bh,Voloshin:2001xi} (see also 
\cite{Leibovich:2002ys}, where a 6--8\% correction on $|V_{ub}|$ was
obtained for $E_0=2.2$\,GeV using a simple model for spectator 
effects). In these references, several strategies have been developed 
that could help to determine the magnitude of these corrections using 
experimental data.

\vspace{0.3cm}  
{\it Acknowledgment:\/} 
This research was supported by the National Science Foundation under 
Grant PHY-0098631. I am grateful to Ed Thorndike and Misha Voloshin 
for useful discussions.

\end{document}